\documentstyle[preprint,aps,amsfonts]{revtex}
\begin{document}

\def\db{$\beta\beta_{0\nu}$\ }
\def\bb{($\beta\beta_{0\nu}$)}

\title{ \small \bf  SIGNALS OF HEAVY MAJORANA NEUTRINOS AT HADRON COLLIDERS}

\author{{\scshape O.~Panella}\footnote{Author to whom correspondence should be addressed. E-mail:
Orlando.Panella@PG.infn.it}$^\S$, {\scshape M.~Cannoni}$^{\P\S}$, 
{\scshape C.~Carimalo}$^\ddagger$ 
 and {\scshape Y.~N.~Srivastava}$^{\P\S}$.
}
\address{
$^{\S}$Istituto Nazionale di Fisica Nucleare, Sezione di Perugia,
 Via A.~Pascoli, I-06123 Perugia, Italy
}
\address{
$^\P$Dipartimento di Fisica, Universit\`a degli Studi di Perugia,
Via A. Pascoli, I-06123, Perugia, Italy}
\address{
$^{\ddagger}$Laboratoire de Physique 
Nucl\'eaire et Hautes Energies, IN2P3-CNRS\\ 
Universit\'e Pierre et Marie Curie, 
4 place Jussieu, Tour 33, F-75252, Paris Cedex 05, France\\
}
\date{\today}

\maketitle
\draft
\begin{abstract}
{The lepton number violating signal of like-sign-dileptons (LSD),
\(p p \to \ell^\pm\ell^\pm + 2\hbox{jets} \),
is investigated within a model of mixing in the neutrino sector assuming
the existence of heavy Majorana neutrino 
states with a left--handed coupling to the light leptons. The LSD signal
receives contributions both from the resonant production of a heavy Majorana
neutrino ($N$) and from the exchange of a virtual $N$
in the WW fusion mechanisms. These two possibilities are discussed in
detail and compared. Helicity amplitudes are given pointing out differences 
with calculations previously reported by other authors. 
The signal cross--sections  are computed at the energy
of the LHC collider (\(\sqrt{S}=14\) TeV) at CERN and within the 
existing experimental limits on the mixing couplings, 
including those coming from neutrinoless double beta 
decay. Detailed angular distributions of signal reactions which are 
complementary to  previous studies on the argument are presented.} 
\end{abstract}


\section{Introduction}
\label{intro}
The quest for a complete understanding of the true nature of neutrino masses
and mixing in the neutrino sector of the Standard Model (SM) is one of the 
most important challenges
that physicists have to face in the coming years.
Perhaps the experimental difficulties involved in the study of
neutrino properties  can be best appreciated noticing that only very recently
the neutral partner of the tau lepton, the tau neutrino ($\nu_{\tau}$),
has been directly observed~\cite{Baller:2001ix}.
In addition, after decades of experimental efforts there is finally some
evidence for neutrino oscillations~\cite{superk} and thus the first
indications of the fact that neutrinos might not be massless particles
as advocated by the SM.  Even so, it is not clear at present
whether these neutral fermions are of Dirac or Majorana type (for a Dirac
neutral $\nu\neq\bar{\nu}$ while for a Majorana $\nu=\bar{\nu}$).
It is well known that if neutrinos are of the Majorana type then the Lepton
number (L) is not conserved and processes which violate L by two units
$\Delta \hbox{L} =\pm 2$ become possible. Perhaps one of the most interesting
lepton number violating processes is the neutrinoless double beta decay
$\beta\beta_{0\nu}$ which currently provides the most stringent bound
on the {\it effective  light} Majorana mass
\(\langle m_\nu \rangle  \leq 0.1\ \hbox{eV}\). Extensive literature is
available on this subject both from the experimental and theoretical sides.
In view of the forthcoming large hadron collider (LHC) to be built at CERN
which will realize proton--proton collisions at energies of
$\sqrt{S} =14\ \hbox{TeV}$ it will be of increasing importance to study
quark reactions which are the high--energy analogues of the  $\beta\beta_{0\nu}$.
This was realized already some time ago and has been the object of several
studies of the signal (lepton number violating) reaction:
\begin{equation}
\label{process4o}
p p \to \ell^\pm \ell^\pm + 2\ \hbox{jets}  \qquad (\Delta \hbox{L} =\pm 2).
\end{equation}
Any model with lepton number violation
(i.e. due to the presence of massive Majorana neutrinos) will provide a non 
zero amplitude for the process
in Eq.~\ref{process4o},  which receives contributions from parton scattering
as depicted in Fig.~\ref{wwfusion} and Fig.~\ref{qqbaranni}.
In a series of recent 
papers~\cite{op1,op2,op3,op4},
the present authors have addressed the question
of lepton number violation within a composite model scenario where the
partner of the light neutrino, the so called excited neutrino, $N$, is assumed
to be of a Majorana type. The phenomenological implications  of this idea
have been investigated in detail both with respect to the low--energy \db
decay and to the high--energy LSD signal in hadron collisions.
In particular, in \cite{op4} we developed a helicity amplitude 
method to efficiently compute all $2 \to 4$ parton subprocesses that contribute
to the reaction in Eq.~(\ref{process4o}).
The cross--sections were estimated to lead, within some regions of the
parameter space, to an observable signal. One particular advantage of the
approach developed in~\cite{op4} is that it gives the possibility
to treat coherently the two mechanisms of WW--fusion and resonant production
(annihilation channel).
Clearly such an approach can be extended to explore the phenomenology
of other models that entail L violation.
In the event of observation
of such a signal, it will be important to know
the contribution to it of the various models beyond the SM and ultimately,
a way to distinguish among them to understand the 
responsible mechanism .

In this work, the model of heavy isosinglets Majorana states that mix
with ordinary light lepton via a left-handed coupling suppressed by a
mixing coefficient, is assumed and studied in detail. While completing this
work a report~\cite{almeida} appeared in the literature which 
also commented upon the same model of heavy
Majorana neutrinos by studying the resonant production at LHC via the
third order process
\begin{equation}
\label{process3o}
p p \to \ell^+ \ell^+ + W^-  \qquad \Delta \hbox{L} = +2
\end{equation}
and arriving at somewhat optimistic results as to the possibility of
observing the signal. 
The process in Eq.~(\ref{process3o}) is described by the Feynman
diagram obtained by that of Fig.~\ref{qqbaranni} but without including the 
hadronic decay products of the final {\em on-shell} W. However, the 
method used here of evaluating analytically all the relevant amplitudes, 
as developed in~\cite{op4},
is totally independent of that of the authors 
of~\cite{almeida} which relies
on the  {\scshape CompHEP}~\cite{comphep} package.
The rest of the paper is organized as follows:
section II gives a description of
the theoretical framework; in section III the amplitudes of the contributing
subprocesses are presented;  in section IV an exact analytic result is
provided for the process in Eq.~\ref{process3o} highlighting differences
with the formula reported in~\cite{almeida} and discussing thoroughly 
how the conclusions reported there are affected;
section V presents some discussion of kinematic cuts and gives some details
of the numerical calculations along with the results; finally section VI 
provides concluding remarks.
\section{THE MODEL}
The mixing of heavy and light neutrinos is a common feature of many
theoretical models beyond the SM.
Heavy Majorana neutral leptons have been well studied  in the literature 
in particular from a model which is the SM augmented with
additional right--handed
neutrino fields. In this scenario, that
can be seen as a low-energy manifestation of $SU(10)$ GUT
theories~\cite{buc91,heusch}, 
one has the following mass term for neutrinos:
\begin{eqnarray}
{\cal L}_{mass}=-\frac{1}{2}
\left(
\begin{array} {cc}
{\overline{\nu}}_{L} & \overline{({\nu}_{R})^{c}}
\end{array}
\right)
{\cal M}\left(
\begin{array} {c} ({\nu}_{L})^{c}\\ {{\nu}_{R}}
\end{array}
\right) +h.c.\;,
\end{eqnarray}
where $({\nu}_{L})^{c}$ and ${{\nu}_{R}}$ are vectors in generation space
with components given by three chiral fields:
\begin{eqnarray} ({\nu}_{L})^{c}=
\left(
\begin{array}{c}
(\Psi^{\nu_e}_L)^{c}\\
(\Psi^{\nu_\mu}_L)^{c}\\
(\Psi^{\nu_\tau}_L)^{c}
\end{array}
\right)\;\;\;\;{{\nu}_{R}}=
\left(
\begin{array}{c}
\Psi^{\nu_e}_R\\\Psi^{\nu_\mu}_R\\\Psi^{\nu_\tau}_R
\end{array}
\right),
\end{eqnarray}
and a mass matrix given by
\begin{eqnarray}
{\cal M}=\left(\begin{array} {cc} 0 & M_{D}\\M_{D}^{T} &
M_{R}\end{array}\right).
\label{massmatrix}
\end{eqnarray}
In general an arbitrary number of right--handed neutrinos can be added,
but for definiteness and simplicity
the most symmetrical case is considered (i.e. that of three 
right-handed fields). The mass
matrix is then $6{\times}6$, with $M_D$ the $3{\times}3$ matrix of Dirac
type masses for neutrinos that is built via the  spontaneous symmetry
breaking and Yukawa couplings as for charged leptons and quarks, while
$M_R$ is the Majorana $3{\times}3$ mass matrix for singlet right--handed 
fields. If one requires the eigenvalues of $M_R$ to be much bigger than those 
of $M_D$, diagonalizing the mass term one has a three generation
``see-saw''~\cite{gell,mohasenj} model, 
with three light and three heavy Majorana
mass eigenstates. Weak interactions produce weak eigenstates that are
superpositions of mass eigenstates through the mixing matrix.
As opposed to the naive one family ``see--saw'' mechanism, where
$\theta\ {\simeq}\ \sqrt{{m_{\nu}}/{M_N}}$, with three generations (or more) of
right--handed neutrino fields, it is
possible to decouple the magnitude of the mixing coefficient of heavy-light
states from the mass eigenvalues of $M_R$,
and the mixing angles can be treated as
phenomenological parameters bounded only by  existing experimental data.
Examples of mass matrices that satisfy this request are given in
ref.~\cite{pil92,buc90,buc91,chang}. In this context, 
lepton number violation in
\(p\-p\) collisions was analyzed in~\cite{dicus,pil92,dattagu}.
Some authors refer instead to
other theoretical scenarios \cite{almeida,gluza} 
derived from super-string
inspired $E_6$ models. Here global symmetries imposed on
the mass matrix decouple the mixing angles from any relation with
masses eigenvalues, so that they are free parameters that can be large (as
in multi-generational ``see--saw'' type
models). Some caution however must be paid in deriving heavy Majorana
masses in these models where every family of the SM is enlarged
both with a right-handed field and a singlet field $S_L$
in the neutrino sector.
The
mass term for neutrinos in the Lagrangian 
is~\cite{mohava,berna,tomma}:
\begin{eqnarray}
{\cal L}_{mass}=-\frac{1}{2}\left(\begin{array} {ccc} {\overline{\nu}}_{L} & \overline{({\nu}_{R})^{c}} & {\bar{S}}_{L}
\end{array}\right){\cal M}\left(\begin{array} {c} ({\nu}_{L})^{c}\\ {\nu}_{R}\\S^{c}_{L} \end{array}\right) +h.c.\;,
\end{eqnarray}
where the mass matrix is given by:
\begin{eqnarray}
{\cal M}=\left(\begin{array} {ccc} 0 & D & 0\\ D^T & 0 & M^T\\ 0 & M & 0
\end{array}\right),
\label{mass99}
\end{eqnarray}
where again the fields represent a collection of three chiral fields
and the block $M$ and $D$ are Dirac type sub-matrices. The zeros are
consequence of the global lepton number conservation imposed on the
mass matrix. The spectrum that one obtains through diagonalization is
three massless neutrinos and three heavy Dirac neutral
leptons~\cite{melo}. Therefore these type of models
cannot be the base for the discussion of the lepton
violating processes that are the object of this work. One could add
a Majorana mass matrix, $\mu$ for
$S_L$~\cite{gonzalez} (lower left \(3\times 3\) block in the mass
matrix of Eq.~(\ref{mass99})), and obtain
a small Majorana mass for the light
neutrinos, but the heavy states remain Dirac particles. Thus in this
class of models it is not very easy to obtain heavy Majorana masses
(in the TeV range) with substantial heavy-light mixing.

Independent of   theoretical
prejudices, in the following the general interaction
Lagrangian between charged leptons, neutrinos, and gauge vector bosons of the
SM
is given within  this class of models where the weak eigenstate neutrinos
are combinations through the mixing matrix of light and heavy {\em physical}
neutrinos. So the latter have the usual left-handed coupling with
leptons with a coupling constant that is the SM weak coupling $g$
multiplied by the appropriate element of the mixing matrix.
The charged and neutral current interaction
Lagrangian densities are given by:
\begin{eqnarray}
{\cal L}_{int}^{cc}&=&-\frac{g}{\sqrt{2}}\sum_{\ell=e,\mu,\tau}
\left\{\sum_{i}{\overline{\Psi}}_{\ell}{\gamma}^{\mu}\frac{1-{\gamma}^{5}}{2}
(B_{\ell\nu})_{{\ell}i}{\Psi}_{{\nu}i}\right. \nonumber \\
&+&\left.\sum_{j}{\overline{\Psi}}_{\ell}{\gamma}^{\mu}
\frac{1-{\gamma}^{5}}{2}(B_{\ell N})_{{\ell}j}{\Psi}_{{N}_j}\right\}
W_\mu^- +h.c.\nonumber \\
{\cal L}_{int}^{nc}&=&-\frac{g'}{2\sin{\theta}_{W}}Z^{\mu}
\sum_{i,j}{\overline{\Psi}}_{{\nu}_{i}}
{\gamma}_{\mu}\frac{1-{\gamma}^{5}}{2}C_{ij}{\Psi}_{{\nu}_{j}}
\label{leff}
\end{eqnarray}
where $B_{\ell n}$ is a $3\times 6$ rectangular mixing matrix that can be
written $B_{\ell n}=\left(B_{\ell\nu}, B_{\ell N}\right)$, with $B_{\ell\nu}$ and
$B_{\ell N}$ the two \(3\times 3\) block matrices
that give the light-light and light-heavy mixing,  while
the  $6\times6$ matrix $C_{ij}$ is given by $C=B^{\dag}B$.  The
properties of $B$ and $C$ are extensively discussed in the
literature~\cite{schec,pirogov,ilpil}.  
We now discuss
the known experimental constraints on mixing angles.
As these heavy states have not yet been observed,
light-heavy mixing has to be inferred by
low--energy phenomenology.
In~\cite{nardi}, a global fit is  performed
on LEP data identifying the following effective mixing
angles:
\begin{eqnarray}
c^{2}_{{\nu}_{\ell}}&=&(B_{\ell \nu}^{\dag}B_{\ell \nu})_{\ell\ell}=\sum_{i}|(B_{\ell \nu})_{i\ell}|^{2}{\equiv}{\cos^{2}}{\theta}_{{\nu}_{\ell}},\nonumber \\
s^{2}_{{\nu}_{\ell}}&=&(B_{\ell N}B_{\ell N}^{\dag})_{jj}=\sum_{j}|(B_{\ell N})_{j}|^{2}{\equiv}\sin^{2}
{\theta}_{{\nu}_{\ell}}\nonumber \\
&=&1-c^{2}_{{\nu}_{\ell}}
=1-\sum_{i=1}^{3}|B_{l_{{\nu}_{i}}}|^{2}=\sum_{j=4}^{6}|B_{l_{{N}_{j}}}|^{2},
\label{effmix} 
\end{eqnarray}
that are bounded from above as:
\begin{equation}
\label{se}
\begin{array}{ccc}
s^{2}_{{\nu}_{e}} < 0.005\phantom{xxxx}&
s^{2}_{{\nu}_{\mu}} < 0.002\phantom{xxxx}&
s^{2}_{{\nu}_{\tau}} < 0.01.
\end{array}
\end{equation}
The authors of ref.~\cite{almeida} have performed an updated analysis
finding roughly the same value for the bound on the electron parameter
\(s^{2}_{{\nu}_{e}} < 0.0052\) but a somewhat more stringent
bound on the  muon  mixing parameter \( s^{2}_{{\nu}_{\mu}} < 0.0001 \),
thereby making the electron LSD channel the most interesting. However for 
first generation leptons it is necessary to consider bounds coming from non 
observation of neutrinoless double beta decay (see below).
In the next sections
calculations are reported for the first family ($\ell = e$) LSD signal
under the assumption that: $i$) only one
heavy Majorana eigenstate is exchanged; $ii$) the maximum allowed value
for the mixing  can be used for the relative element of the
mixing matrix, i.e. $|B_{eN}|^{2}=0.0052$.  
It should be recalled that when  extensions of the SM
are embedded in more general theories, one  needs to consider
new gauge bosons that can interact with the heavy neutrino states.
However, as these bosons are predicted to be very heavy, 
their interactions are suppressed relative to those of the SM   so that
one can safely neglect them~\cite{almeida,buc91} in the following discussion.

\subsection{Unitarity and \db constraints } 
Before entering into the detailed calculations, a comment is in order,  
regarding violation of unitarity,  and the bound coming from the  
non observation of neutrinoless double beta decay 
\bb\ which is quite important.  
It is well known~\cite{dicus,belanger,gluza} 
that the basic process \( e^- e^- \to W^-W^-\) 
or, equivalently, \( W^-W^-\to e^- e^- \) intervening
in the reactions  discussed in the present work (see Fig.~\ref{wwfusion}),
violates unitarity unless the following condition is satisfied:
\begin{equation}
\sum_\nu m_\nu\, (B_{\ell \nu})^2 +\sum_N M_N\, (B_{\ell N})^2 = 0\, .
\end{equation}
In the model considered here this is automatically satisfied as it is 
easily shown that 
\( \sum_\nu m_\nu\, (B_{\ell \nu})^2 + \sum_N M_N\, (B_{\ell N})^2 
= M_{\ell \ell}\)
where \(M_{\ell \ell} \) is the left-Majorana mass matrix (\(3\times 3\) 
block matrix) of \(\nu_\ell\), and this
is zero as can be inferred from 
the mass matrix given in Eq.~(\ref{massmatrix}).
The Majorana mass term of \(\nu_\ell\) could be different from zero
within a Higgs triplet model which however has been excluded by 
the precision LEP data,  unless one 
considers somewhat more contrived models, where,
for example, the Higgs triplet  is mixed with a singlet allowing to evade 
the LEP bounds~\cite{choi}. 
In the following such eventualities are not considered, 
the absence of a Higgs triplet is assumed, ensuring thus the
unitarity condition to be satisfied according to Eq.~(\ref{massmatrix}).

The non observation of the neutrinoless double beta decay (a second order 
weak nuclear decay) gives the following  bound on the 
{\em effective inverse mass}~\cite{belanger}:
\begin{eqnarray}
\left |\, \sum_{N}\frac{B_{eN}^{2}}{M_N} \, \right|\;<\;
7{\times}\;{10^{-5}}\;\;{\hbox{TeV}}^{-1},
\label{boundbb}
\end{eqnarray}
Within the rather reasonable hypothesis that only the light mass 
eigenstate contributes significantly to both the low-energy \bb\ process 
and high-energy (LSD) signal and using 
\(|B_{eN}|^2\, {\simeq}\, 0.0052\) 
one easily finds a rather impressive lower bound on the mass~\cite{belanger}: 
\begin{equation} 
M_N > 75\, \hbox{TeV} .
\end{equation}
This of course would preclude the possibility of producing such particles
at any forthcoming hadron or lepton collider.
The \db bound could obviously be evaded  as the mixing 
coefficients appear as squared and not through their moduli squared in 
Eq~(\ref{boundbb}). 
Indeed in ref.~\cite{belanger}
several scenarios of mixing are discussed 
where cancellations between different terms arise thus evading the 
\db bound.
Explicit examples are provided~\cite{belanger} 
but of course the resulting models appear somewhat fine tuned and/or 
too much contrived and un-natural.
It would then appear that a phenomenological analysis of the processes
discussed in the introduction would be pointless for heavy neutrino masses
within the mass range \( 100 \hbox{ GeV} \leq M_N  \leq 1 \hbox{ TeV}\).
However, one can 
build models which do not suffer such drawbacks of fine tuning and 
unnaturalness by simply assuming that the heavy Majorana neutrinos do not have
the same  CP parities (\(\eta_{CP}\)). This was done by the authors of 
ref.~\cite{gluza} who considered a model with two degenerate
heavy Majorana neutrinos of \(M_N\approx 1\) TeV and opposite CP parities
that evades the \db bound~\cite{gluza}.
It is within such a scenario that the following phenomenological
study of the processes in Eq.~(\ref{process3o}) and Eq.~(\ref{process4o})
has to be understood.

In any case it must be kept in mind that 
the predictions reported here are based on the maximum 
experimentally allowed mixing and therefore new global fits on the last 
LEP data may lower its value, and thus affect the present conclusions.

\section{AMPLITUDES OF L-VIOLATING PARTON SUBPROCESS}
In the following,  the helicity amplitudes for parton
sub-processes that contribute to production of LSD via the exchange
(or production) of a heavy Majorana neutrino are presented.
The effective interaction used is that of Eq.~(\ref{leff}).
Considering for the moment only the first fami\-ly,
three
different types of processes should be distinguished:
\(
(i)\,
u u     \to d d     + \ell^+\ell^+, \)
\(
(ii)\,  u \bar{d} \to d \bar{u} + \ell^+\ell^+, \)
\(
(iii) \,
\bar{d}   \bar{d}   \to \bar{u}   \bar{u}   + \ell^+\ell^+.
\)

Let the tensor $T_{\mu\nu}$  describe the virtual sub-process
$W^\star W^\star \to \ell^+\ell^+$ (Fig.~\ref{wwfusion}),
while the tensor $\widetilde{T}_{\mu\nu}$
describes the virtual sub-process
$(W^{\star})^+ \to \ell^+\ell^+ (W^{\star})^- $ appearing
in the diagram of Fig.~\ref{qqbaranni}.
$J_{a,c}$ and $\bar{J}_{b,d}$, are the quark (anti-quark)
currents that couple in the t-channel to the
virtual gauge bosons of the standard model (Fig.~\ref{wwfusion}) while
$\widetilde{J}_{a,b}$ and $\widetilde{J}_{c,d}^* $,
are the incoming and outgoing currents of the $q\bar{q}'$ pair
that couples in the s-channel to the W-bosons (Fig.~\ref{qqbaranni}):
\begin{equation}
\begin{array}{ll}
J^\mu_{a,c}
 =  \bar{u}(p_c)\, \gamma^\mu\, P_L\, u(p_a)\,
&\bar{J}^\mu_{b,d}
 =  \bar{v}(p_b)\, \gamma^\mu\, P_L\, v(p_d)\, \\
\widetilde{J}^\mu_{a,b}
 =  \bar{v}(p_b)\, \gamma^\mu\, P_L\, u(p_a)\,
&({\widetilde{J}^{\mu}_{c,d}})^*
 =  \bar{u}(p_c)\, \gamma^\mu\, P_L\, v(p_d)\
\end{array}
\end{equation}
with $P_L = (1-\gamma_5)/{2}$. 

With the following definitions of propagator
factors:
\begin{eqnarray}
1/A &=& \left[(p_a -p_c)^2-M_W^2\right]
      \left[(p_b -p_d)^2-M_W^2\right]\cr
1/B &=& \left[(p_a -p_d)^2-M_W^2\right]
      \left[(p_b -p_c)^2-M_W^2\right]\cr
1/\widetilde{A} &=& \left[(p_a +p_b)^2-M_W^2 +iM_W\Gamma_W\right]
      \left[(p_c +p_d)^2-M_W^2+iM_W\Gamma_W\right]
\end{eqnarray}
\begin{equation}
\label{propfac}
\begin{array}{ll}
C = {(p_a -p_c-p_e)^2-M_N^2}\phantom{+iM_N\Gamma_N} &
D = {(p_a -p_c-p_f)^2-M_N^2}\phantom{+iM_N\Gamma_N} \\
E = {(p_a -p_d-p_e)^2-M_N^2}\phantom{+iM_N\Gamma_N} &
F = {(p_a -p_d-p_f)^2-M_N^2}\phantom{+iM_N\Gamma_N} \\
\widetilde{C} = {(p_a +p_b-p_e)^2-M_N^2+iM_N\Gamma_N} &
\widetilde{D} = {(p_a +p_b-p_f)^2-M_N^2+iM_N\Gamma_N}
\end{array}
\end{equation}
the amplitudes (in the unitary gauge) are :

($i$)
\underline{{$U_i U_j \to D_k D_l + \ell^+\ell^+$}}
\begin{equation}
{\cal M} = {\cal K} \,
\left\{ \, V^*_{U_iD_k} V^*_{U_jD_l} \, A \,  \left[ J^\mu_{(a,c)} \,   T_{\mu\nu}\,
\, {J}^\nu_{(b,d)}\, \right]
- V^*_{U_iD_l} V^*_{U_jD_k} \, B \,
\Big[\, (p_c \leftrightarrow p_d)\, \Big] \right\}  \,  ;
\end{equation}

$(ii)$
\underline{\  $U_i \bar{D}_j   \to {D}_k   \bar{U}_l
+ \ell^+\ell^+ \,$}
\begin{eqnarray}
\label{ampiiannihi}
{\cal M}{(WW-\hbox{fusion})}& = &{\cal K} \, (V_{U_iD_k})^* (V_{U_lD_j})^* \,
 \left[ \, A\, J^\mu_{(a,c)} \,  T_{\mu\nu}\,  \bar{J}^\nu_{(b,d)}\,
\right] \,   ;\cr
{\cal M}{(q\bar{q}'-\hbox{annihilation})}
& = &{\cal K} \, (V_{U_iD_j})^* (V_{U_lD_k})^* \,
 \left[ \, \widetilde{A}\, \widetilde{J}^\mu_{(a,b)} \,
\widetilde{T}_{\mu\nu}\,  (\widetilde{J}^\nu_{(c,d)})^*\,
\right] \,  ;
\end{eqnarray}

($iii$)
\underline{{$\bar{D}_i \bar{D}_j \to \bar{U}_k \bar{U}_l
+ \ell^+\ell^+$}}
\begin{equation}
{\cal M} = {\cal K} \,
\left\{ \, V_{U_kD_i}^* V_{U_lD_j}^* \, A \,
\left[ \bar{J}^\mu_{(a,c)} \,   T_{\mu\nu}\,
\, \bar{J}^\nu_{(b,d)}\, \right] -
V_{U_lD_i}^* V_{U_kD_j}^* \, B
\Big[ \, ( p_c \leftrightarrow p_d )\, \Big] \right\} \,  ;
\end{equation}
where $U_i$ denotes a positively charged quark (up-type) while $D_i$ denotes
a negatively charged one (down-type).
The quantities $V_{U_iD_j}$ are the elements
of the CKM mixing matrix. Of course the annihilation diagram of Fig. 2 comes in
only in quark-anti-quark scattering.
In processes ($i$) and ($iii$) the part of the amplitude depending on the
factor $B$ is due to the diagrams obtained upon exchanging the
final state quarks.
In the framework of the Lagrangian
c.f. Eq.~(\ref{leff}), as discussed in section II,
we have:
\begin{eqnarray}
T_{\mu\nu} & = & \bar{u}(p_e)\,
\left[ \frac{\gamma_\mu\gamma_\nu}{C}
+ \frac{\gamma_\nu\gamma_\mu}{D}
\right] \frac{1-\gamma_5}{2} \,\, v(p_f) \, ,
\cr
\widetilde{T}_{\mu\nu} & = &  \bar{u}(p_e)\,
\left[ \frac{\gamma_\mu\gamma_\nu}
{\widetilde{C}}
+ \frac{\gamma_\nu\gamma_\mu}{\widetilde{D}}
\right] \frac{1-\gamma_5}{2} \,\, v(p_f) \, ,
\cr
{\cal  K}  &= & \frac{g^4}{4} \left( B_{eN} \right)^2
\, M_N\, .
\label{Tmunutensor}
\end{eqnarray}
Due to the chiral nature of the couplings involved,
the calculation is particularly simple if performed in
the helicity basis~\cite{kleiss}. Within this approach the amplitudes
are evaluated in terms of {\em scalar spinor  products}
instead of scalar products of
particle's momenta as in the usual method of squaring the amplitudes.
In the massless approximation (all $p_i^2 = 0 $ for external particles)
the only non zero scalar products between spinors of given helicity
are the quantities:
\begin{eqnarray}
s(m,n) &=& s(p_m,p_n) = \bar{u}_+(p_m) u_-(p_n)\, ,\nonumber \\
t(m,n) &=& t(p_m,p_n) = \bar{u}_-(p_m) u_+(p_n)\, ,
\end{eqnarray}
which are given by:
\begin{eqnarray}
s(m,n) & = & - 2\sqrt{E_mE_n}\ \  G_{mn}, \nonumber \\
t(m,n) & = & + 2\sqrt{E_mE_n}\ \  F_{mn},
\label{su}
\end{eqnarray}
with:
\begin{eqnarray}
G_{mn} & = & \cos(\theta_m/2)\sin(\theta_n/2)
\, e^{ +i(\phi_m-\phi_n)/2 }
-\sin(\theta_m/2)\cos(\theta_n/2)
\, e^{ -i(\phi_m-\phi_n)/2} , \nonumber \\
F_{mn} & = &  (G_{mn})^*.
\label{gf}
\end{eqnarray}
Consider for a moment the lepton line and its corresponding tensor
$T_{\mu\nu}$ projected out on the helicity basis:
\begin{equation}
T_{\mu\nu}^{(\lambda,\lambda')}(p_e,p_f) = \bar{u}_{\lambda}\, (p_e)\,
\left[ \frac{\gamma_\mu\gamma_\nu}{C}
+ \frac{\gamma_\nu\gamma_\mu}{D}
\right] \frac{1-\gamma_5}{2} \,\, v_{\lambda'}\,(p_f)
\end{equation}
It is easily shown (use $v_{\lambda}\,(p) = -2\lambda \gamma_5
u_{-\lambda}(p)$)
that only one of the four helicity combinations is non
vanishing:
\begin{equation}
\begin{array}{ll}
T_{\mu\nu}^{(+,+)}(p_e,p_f) =  0 \qquad&
T_{\mu\nu}^{(-,-)}(p_e,p_f) =  0\cr
T_{\mu\nu}^{(-,+)}(p_e,p_f) =  0 \qquad&
T_{\mu\nu}^{(+,-)}(p_e,p_f) =  \bar{u}_{+}\, (p_e)\,
\left[ \frac{\gamma_\mu\gamma_\nu}{C}
+ \frac{\gamma_\nu\gamma_\mu}{D}
\right] \, u_{-}\,(p_f)
\end{array}
\end{equation}
In particular it follows that if $p_e=p_f = p $ then $C=D$ and:
\begin{equation}
T_{\mu\nu}^{(+,-)}(p,p) = 2\, \frac{\eta_{\mu\nu}}{C}\, \bar{u}_+(p)u_-(p)
= s(p,p)=0.
\end{equation}
Due to the structure of the coupling the two massless  (identical)
leptons have the same helicity and therefore by the Pauli exclusion principle
if they have the same momentum the amplitude must vanish. All amplitudes given
below satisfy this property.
The above remarks apply as well to the quark lines
and in the massless approximation
only one helicity amplitude is non zero.
The following result is found:

\noindent ($i$) \underline{{$U_i U_j \to D_k D_l + \ell^+\ell^+$}}
\begin{eqnarray}
{\cal M}= 4\,  {\cal K} \, t(c,d) & \Bigg\{ &
\phantom{+} (V_{U_iD_k})^* (V_{U_jD_l})^* \, A \,\,
\left[\frac{ s(e,a) s(b,f)}{C}-\frac{s(f,a)s(b,e)}{D}
\right] \nonumber \\
& \, & + (V_{U_iD_l})^* (V_{U_jD_k})^* \, B \,\,
\left[\frac{s(e,a) s(b,f)}{E}-\frac{s(f,a)s(b,e)}{F}
\right]\Bigg\},
\label{ampsi}
\end{eqnarray}

\noindent ($ii$)
\underline{{$U_i \bar{D}_j \to D_k \bar{U}_l + \ell^+\ell^+$}}
\begin{eqnarray}
{\cal M}= 4\, {\cal K}\, t(c,b)\,
&\Bigg\{ &\phantom{+} (V_{U_iD_k})^* (V_{U_lD_j})^* \,  A\,
\left[\frac{ s(e,a) s(d,f)}{C}-\frac{s(f,a)s(d,e)}{D}
\right]    \nonumber \\
&\, & + (V_{U_iD_j})^* (V_{U_lD_k})^* \, \tilde{A}\,
\left[\frac{ s(e,a) s(d,f)}{\tilde{C}}-\frac{s(f,a)s(d,e)}{\tilde{D}}
\right] \Bigg\}      ,
\label{ampii}
\end{eqnarray}

\noindent ($iii$) \underline{{$\bar{D}_i \bar{D}_j \to \bar{U}_k \bar{U}_l
+ \ell^+\ell^+$}}
\begin{eqnarray}
{\cal M}= 4\,  {\cal K} \, t(a,b) & \Bigg\{ &
\phantom{+} (V_{U_kD_i})^* (V_{U_lD_j})^*\, A \,\,
\left[\frac{ s(e,c) s(d,f)}{C}-\frac{s(f,c)s(d,e)}{D}
\right] \nonumber \\
& \, & -(V_{U_lD_i})^* (V_{U_kD_j})^*\, B \,\,
\left[\frac{s(e,d) s(c,f)}{E}-\frac{s(f,d)s(c,e)}{F}
\right]\Bigg\}.
\end{eqnarray}
The above simple analytic form of the amplitudes is also quite easy
to implement in a code for numerical applications, since the quantities
$s(p_i,p_j)$ and  $t(p_i,p_j)$ are just functions of the energies and angles
of the particle's momenta, c.f. Equations~(\ref{su}) and (\ref{gf}).

In addition it is also possible to derive from the above expression of
the amplitudes the average squared matrix elements for the various processes
in terms of scalar products of particle's four momenta.
One finds using $|s(p_i,p_j)|^{2}=|t(p_i,p_j)|^{2}=2p_i{\cdot}p_j$:

\noindent ($i$) \underline{{$U_i U_j \to D_k D_l + \ell^+\ell^+$}}
\begin{eqnarray}
\sum|{\cal M}|^{2}&=&128\, {\cal K}^{2}\,|V_{U_iD_k}^*V_{U_jD_l}^*|^2 \,
p_c{\cdot}p_{d}\,
\Bigg\{p_{a}{\cdot}p_{e}p_{b}{\cdot}p_{f}
\left|\frac{A}{C}+\xi_1^*\frac{B}{E}\right|^{2}
+p_{a}{\cdot}p_{f}p_{b}{\cdot}p_{e}
\left|\frac{A}{D}+\xi_1^*\frac{B}{F}\right|^{2}
\cr
&&\phantom{xxxxxxxxxxxxxxxxxxxx}- L(p_a,p_f,p_b,p_e) \,
\Re\, X_1  +  \varepsilon(p_a,p_f,p_b,p_e)\, {\Im}\, X_1 \Bigg\}
\label{quad1}
\end{eqnarray}
\noindent ($ii$) \underline{{$U_i \bar{D}_j \to D_k \bar{U}_l + \ell^+\ell^+$}}
\begin{eqnarray}
\sum|{\cal M}|^{2}&=&128\, {\cal K}^{2}\,|V_{U_iD_k}^*V_{U_lD_j}^*|^2 \,
p_c{\cdot}p_{b}\,
\Bigg\{p_{a}{\cdot}p_{e}p_{d}{\cdot}p_{f}
\left|\frac{A}{C}+\xi_2^*\frac{\widetilde{A}}{\widetilde{C}}\right|^{2}
+p_{a}{\cdot}p_{f}p_{d}{\cdot}p_{e}
\left|\frac{A}{D}+\xi_2^*\frac{\widetilde{A}}{\widetilde{D}}\right|^{2}\cr
&&\phantom{xxxxxxxxxxxxxxxxxxxx} - L(p_a,p_f,p_d,p_e) \,
\Re\, X_2 +\varepsilon(p_a,p_f,p_d,p_e)\, {\Im}\, X_2\Bigg\}
\label{quad2}
\end{eqnarray}
\noindent ($iii$) \underline{{$\bar{D}_i \bar{D}_j \to \bar{U}_k \bar{U}_l
+ \ell^+\ell^+$}}
\begin{eqnarray}
\sum|{\cal M}|^{2}&=&128\, {\cal K}^{2}\,|V_{U_kD_i}^*V_{U_lD_j}^*|^2 \,
p_a{\cdot}p_{b}\,
\Bigg\{p_{c}{\cdot}p_{e}p_{d}{\cdot}p_{f}
\left|\frac{A}{C}+\xi_3^*\frac{B}{F}\right|^{2}
+p_{c}{\cdot}p_{f}p_{d}{\cdot}p_{e}\left|\frac{A}{D}
+\xi_3^*\frac{B}{E}\right|^{2}
\cr
&&\phantom{xxxxxxxxxxxxxxxxxxxx}- L(p_c,p_f,p_d,p_e) \,
\Re\, X_3  +  \varepsilon(p_c,p_f,p_d,p_e)\, {\Im}\, X_3 \Bigg\}
\label{quad3}
\end{eqnarray}
where the following definitions have been used:
\begin{equation}
\begin{array}{ll}
X_{1}=
\Big(\frac{A}{C}+\xi_{1}^*\frac{B}{E}\Big)\Big(\frac{A}{D}
+\xi^*_{1}\frac{B}{F}\Big)^* & \quad \xi_{1}=
{V_{U_iD_l}V_{U_jD_k}}/({V_{U_iD_k}V_{U_jD_l}})\\
X_{2}
=\left(\frac{A}{C}+\xi_{2}^*\frac{\widetilde{A}}{\widetilde{C}}
\right)
\Big(\frac{A}{D}+\xi^*_{2}\frac{\widetilde{A}}{\widetilde{D}}\Big)^* &
\quad \xi_{2}=
V_{U_iD_j}V_{U_lD_k}/(V_{U_iD_k}V_{U_lD_j})\\
X_3=\Big(\frac{A}{C}+\xi_3^*\frac{B}{F}\Big)
\Big(\frac{A}{D}+\xi_3^*\frac{B}{E}\Big)^* & \quad \xi_3=
V_{U_lD_i}V_{U_kD_j}/(V_{U_kD_i}V_{U_lD_j})
\end{array}
\end{equation}
\begin{eqnarray}
L(p_1,p_2,p_3,p_4)&=&p_1{\cdot}p_2\,p_3{\cdot}p_4-p_1{\cdot}p_3\,p_2{\cdot}p_4
+p_1{\cdot}p_4\,p_2{\cdot}p_3,\cr
\varepsilon (p_1,p_2,p_3,p_4) & = & \varepsilon^{\mu\nu\lambda\rho}
(p_1)_\mu(p_2)_\nu(p_3)_\lambda(p_4)_\rho
\end{eqnarray}
with $\varepsilon^{0123}=+1$.
The above formulas (\ref{quad1}), (\ref{quad2}), (\ref{quad3})
have been written down in all
generality allowing for possibility that the matrix elements of the
CKM matrix might be complex. Except for s-channel contributions, such 
squared matrix elements were already given in \cite{dicus} in the framework 
of the same model as that here used. However, the formulas found in 
\cite{dicus} disagree from ours in various aspect, and in the fact that they 
seem to have been obtained as if a \(({1+\gamma^5})/{2}\) were used for the
coupling 
between leptons, Majorana neutrinos and 
$W^-$, instead of the usual \(({1-\gamma^5})/{2}\) for left--handed coupling 
(formula (2) in \cite{dicus} should then contain a \(({1+\gamma^5})/{2}\) 
factor). The numerical applications
reported here concern only quarks of the first and second family and
{\em for those} the elements of the
CKM matrix can be approximated (Wolfenstein 
parameterization~\cite{wolf})
to a high degree of accuracy to be real.
Accordingly $\xi_1, \xi_2, \xi_3$ will be real and
$\Im X_1 =\Im X_3 = 0$. Only $X_2$ will retain an imaginary part
due to the complex resonant propagator factors.

The decay width of the heavy neutrino, $\Gamma_N$, which appears in the
factors $\widetilde{C}$ and $\widetilde{D}$ is of course
a quantity which depends on the parameters of the particular
model that is being considered here, $B_{eN}$ and  $M_N$.
It receives contributions from the two decay channels $N_\ell \to W+ \ell $
(charged current) $ N_\ell \to Z + \nu_\ell$ (neutral current).
In the present model there are no other decay channels
open. Some authors however~\cite{dattagu,gluza} include also possible Higgs
interactions of the heavy neutrino states. Doing so will open up additional
decay modes depending on the mass relation between $M_N$ and $M_H$.
As both of these parameters are still unknown the present authors prefer
not to consider this possibility. Hence the values used for $\Gamma_N$
in the present estimates of the resonant cross--sections should be regarded
as {\em lower bounds} for the heavy neutrino total width.
As the total decay width appears in the resonant propagator factors,
it is a quite important parameter as regards numerical applications.
Thus in appendix A are reported the formulas used to evaluate the
partial and total decay widths, while in Table~I the corresponding
numerical values of the total
decay width $\Gamma_N$ are reported for different values of the
neutrino mass $M_N$.
It can be inferred that the total decay width changes quite drastically
going from $\Gamma_N/M_N = 0.0011 \% $ at $M_N = 100$ GeV to
$\Gamma_N/M_N = 0.43\, \% $ at $M_N = 1000$ GeV.
In addition it is to be remarked that the authors of~\cite{almeida} 
quote
even smaller values of the total decay width (see Table I) which appear to
be unsubstantiated within the present model. 
Using the interaction of
Eq.~(\ref{leff}) one readily evaluates in the unitary gauge the 
partial widths (assume $M_N \gtrsim 100$ GeV)
$\Gamma (N_l \to W +l)$ and $\Gamma (N_l \to Z +\nu_l)$ and hence the total
decay width.
In the appendix  
it is shown explicitly that the values of the width quoted by the 
authors of ref.~\cite{almeida}
can be reproduced by calculating in the 't Hooft gauge but not including 
the ghosts (Goldstone bosons) diagram.

In reference~\cite{op3} in order to keep the numerical computations
of cross-sections reasonably simple, a constant value of $\Gamma_N$ 
was assumed. Here this is not possible
as the width of the heavy neutrino varies so strongly between
low ($\approx 100$\ GeV) and high ($\approx 1000$\ GeV) masses, and one is
obliged to keep the functional dependence of the total width on
the neutrino mass $M_N$. This makes the contribution
of the resonant diagrams strongly dependent on the heavy neutrino mass $M_N$.

\section{The third order process ${\lowercase{pp} \to {\ell}^+ {\ell}^+ + W^-}$}
This process has been the object of~\cite{almeida} where the square of
the amplitude and the subsequent phase-space integration were calculated
by means of the {\scshape CompHEP} package~\cite{comphep}.
The amplitude of the third order process is:
\begin{eqnarray}
{\cal M} &=&{\cal K}' \, \widetilde{A}' \, \bar{v}(p_b)\,  \gamma^\mu\,
\frac{1-\gamma_5}{2} \,
u(p_a) \, \tilde{T}_{\mu\nu} \, [\epsilon^\nu_{(\lambda)}(Q)]^*\cr
{\cal K}' &=& \left(\frac{g}{\sqrt{2}}\right)^{3}\, V_{ud}^{*} \, B_{eN}^{2}\, M_{N}\cr
1/\widetilde{A}' &=& \left[(p_a +p_b)^2-M_W^2 +iM_W\Gamma_W\right]\cr
\widetilde{C}' &=& (p_e +Q)^2-M_N^2 +iM_N\Gamma_N,	\cr
\widetilde{D}' &=& (p_f +Q)^2-M_N^2 +iM_N\Gamma_N,
\label{eewamp}
\end{eqnarray}
with $\widetilde{T}_{\mu\nu}$  defined in Eq.~(\ref{Tmunutensor}).
Clearly relative to the annihilation amplitude of the process
$ u\, \bar{d} \to
\ell^+ \ell^+ + u\, \bar{d} $, given in Eq.~(\ref{ampiiannihi}),
the outgoing quark's  current $[\widetilde{J}^\nu_{(c,d)}]^*$ is
replaced by the on--shell W gauge boson polarization vector
$[\epsilon^\nu_{(\lambda)}(Q)]^*$ and the constants $\widetilde{A}'$ and
${\cal K}'$ are accordingly redefined.

Here a formula is given which was derived  by means of the standard
method of squaring the amplitude and summing over polarizations
of initial and final particles thus reducing the calculation  to the product
of traces of strings of Dirac matrices.
\begin{eqnarray}
\sum_{\text{spins}}|{\cal M}|^{2}&=&16\, |{\cal K}'\, \tilde{A}'|^{2}\cr
&\times&\left\{\frac{1}{|\tilde{C}'|^{2}}p_{a}{\cdot}p_e
\left(p_{b}{\cdot}p_f+2
\frac{Q{\cdot}p_fQ{\cdot}p_{b}}{M_{W}^{2}}\right)\right.
+\frac{1}{|\tilde{D'}|^{2}}p_{a}{\cdot}p_f\left(p_{b}{\cdot}p_e+2
\frac{Q{\cdot}p_eQ{\cdot}p_{b}}{M_{W}^{2}}\right)\cr
&+&{\Re}e\left(\frac{1}{\tilde{C}'{\tilde{D'}}^{*}}\right)\left[p_e{\cdot}p_f
\left(p_{a}{\cdot}p_{b}
+2\frac{Q{\cdot}p_{a}Q{\cdot}p_{b}}{M_{W}^{2}}\right)\right.
-p_{a}{\cdot}p_f
\left(p_{b}{\cdot}p_e+2\frac{Q{\cdot}p_eQ{\cdot}p_{b}}{M_{W}^{2}}\right)\cr
&-&\left.p_{a}{\cdot}p_e\left(p_{b}{\cdot}p_f
+2\frac{Q{\cdot}p_fQ{\cdot}p_{b}}{M_{W}^{2}}\right)\right]
-\left.{\Im}m \left[
\frac{\varepsilon(p_{a},p_f,p_{b},p_e)}{\tilde{C}'{\tilde{D'}}^{*}}
\right]
\left(1+2\frac{Q{\cdot}p_{b}}{M_{W}^{2}}\right)
\right\},
\label{eew}
\end{eqnarray}
where
$Q$ is the W boson four-momentum.
It should also be remarked that the above result has been verified against
that of the annihilation part of
Eq.~(\ref{ampii}) by taking the small width approximation on the
latter, factorizing the phase-space and integrating out that of the
two quarks from the on-shell W.

However the formula given
by the authors of ref.~\cite{almeida} 
(c.f in the appendix) appears not to
take into account the fact that the amplitude is the sum of
two diagrams antisymmetrized with respect to the
variables of the two identical leptons in the final state. The two diagrams
have different heavy Majorana neutrino propagator factors 
(c.f \(\tilde{C'}\) and \(\tilde{D'}\) in Eq.~(\ref{eewamp})).
Moreover all factors explicitly dependent on  \((M_W^2)^{-1}\) and coming 
from the sum over  the real W polarizations 
 \( \left[\sum_\lambda \epsilon_\lambda(k) \epsilon^*_\lambda(k) = -\eta_{\mu\nu}+
(k_\mu k_\nu)/M_W^2 \right] \) are  missing.
They claim that their result is the output
of the {\scshape CompHEP} software package~\cite{comphep} 
which allowed them to perform
symbolic computation of matrix elements.  
In order to understand the reason of this discrepancy the present 
authors decided to verify again the result given in Eq.~\ref{eew}
by using the same {\scshape CompHEP} software package used by the authors 
of ref.~\cite{almeida}. It appears that in {\scshape CompHEP} non standard models
have to be defined, in the sense that new particles and new 
interaction vertices have to be defined by the user. {\scshape CompHEP} can perform
calculations both in the unitary gauge and in the 't Hooft gauge.
Defining the heavy neutrino and its interactions as in Eq.~(\ref{leff})
and working in the {\em Unitary gauge} it has been verified 
that the {\scshape CompHEP} output reproduces that reported in Eq.~(\ref{eew})
up to the factor containing the totally antisymmetric 
Levi-Civita tensor: this is because the 
program directly replaces the squared propagator for the resonant particles.
However, adding this term to the {\scshape CompHEP} fortran output  of 
the matrix elements, it is found numerically irrelevant as it is proportional
to the width of the heavy neutrino $N$.
It is thus established that  Eq.~(\ref{eew}) 
agrees  with the {\scshape CompHEP} output in the unitary gauge up to terms 
proportional to the width of the resonant particle (neutrino).

Working with {\scshape CompHEP} in the {\em 't Hooft gauge} requires some 
care as
in this gauge also the couplings of the $W^\pm$ and  $Z$ Goldstone bosons 
(ghosts) have to be defined by the user. These are necessary to reproduce 
the correct sum over  polarizations for $W^\pm$ and  $Z$.
With the inclusion of the ghost vertices the calculation in the 't Hooft 
gauge reproduces the result of the Unitary gauge (as it should) which agrees 
with that of Eq.~(\ref{eew}).
Doing so the present authors realized 
that the result of ref.~\cite{almeida} [c.f. Eq.~(A1)] 
is indeed obtained with the 
{\scshape CompHEP} package working in the 't Hooft gauge if one 
{\em does not defines the ghost vertices}. The discrepancy of the result
of this  report, Eq.~(\ref{eew}), 
with that of Eq.(A1) of ref.~\cite{almeida} is thus resolved: 
the formula given in ref.~\cite{almeida} is just one of the terms when 
working  in the 't Hooft gauge; the ghost contribution needs to be added
to get the complete result which agrees with that reported here in 
Eq.~(\ref{eew}) which is obtained in the unitary gauge.   

In  the appendix formulas are presented which illustrate this point explicitly
for the simpler case of the  heavy neutrino 
width. The calculation without including the ghost diagrams amounts to use 
[\( \sum_\lambda \epsilon_\lambda(k) \epsilon^*_\lambda(k) 
= -\eta_{\mu\nu} \)] 
for the sum over the gauge boson polarizations and the width
thus obtained
reproduces the anomalously small values  reported in
ref.~\cite{almeida}
for $M_N=100,\;(800)$ GeV.  

The total cross--section, summing up the contributions of the quarks of the
first and second generations, is given by:
\begin{eqnarray}
\sigma = \frac{1}{2\hat{s}} \, \int \, dx_a\, dx_b \,
\sum_{U\bar{D}} &\,&\left[f_U(x_a)f_{\bar{D}}(x_b)+
x_a \leftrightarrow x_b)\right] \cr
&\,& \times\, \frac{1}{4}\,\times \frac{1}{3}\sum_{\hbox{spins}}\,
|{\cal M}|^2 ( U \bar{D} \to e^+e^+ W)
\, \frac{d_3{\hbox{LIPS}}(\sqrt{\hat{s}};p_e,p_f,Q)}{2}\, ,
\label{sigmatot}
\end{eqnarray}
where the numerical factors account for average on initial spins and colors.
All numerical results presented in the next section will be at the level 
of a parton Monte Carlo, while aspects like jet 
hadronization and detector simulations are beyond the scope of this work.

\section{Discussion and  results}
The previous observation clearly does 
not affect the W-W fusion matrix elements 
because in the t-channel there is no finite width effect and the gauge boson 
propagator's numerator is simply $\eta_{\mu\nu}$ also in unitary gauge as 
the limit of massless external particles is assumed. 
As stated, correctly, in ref.~\cite{almeida} these 
diagrams give rise to a very  small cross--section, 
$\sigma\, {\simeq}\, 10^{-5}-10^{-4}$ fb in 
the range of neutrino masses of interest here 
\(100 \hbox{\ GeV} \leq M_N \leq 1000 \hbox{\ GeV}\). At the LHC assuming
an integrated luminosity over one year of ${{L}}=100$ fb$^{-1}$ 
one could define a minimal cross--section \(\sigma_0=10^{-2}\) fb that 
would give 1 event per year. Clearly, smaller cross--section would be 
impossible to measure (with statistical significance) on a one year run.
Thus, for the model studied in this work, the WW-fusion mechanism is 
negligible and not measurable.
This was  checked :
\((i)\) with authors'  defined fortran code~\cite{op4} based on 
Eqs.~(\ref{quad1},\ref{quad2},\ref{quad3}) and feeding them into 
the phase space integration routines provided within the {\scshape CompHEP} package; 
\((ii)\) and directly with {\scshape CompHEP} performing a 
symbolic matrix element calculation followed by numerical ({\scshape VEGAS})
phase space integration. 

Therefore the following discussion will concentrate on the resonant 
production mechanism of Fig.~\ref{qqbaranni}, with a particular emphasis on 
the third order process and the comparison with the result 
reported in~\cite{almeida}.
The following  numerical results were obtained with {\scshape CompHEP}, using the 
same  kinematic 
cuts of ref.~\cite{almeida}: 
$p_T^{lepton}>5$ GeV, $p_T^W>15$ GeV and $|\eta|<2.5$ 
for the final state particles. The parton distributions 
functions (PDF) used are the set of MRS[G]~\cite{mrs}. As shown explicitly 
in~\cite{op4} these type of calculations are not very sensitive to the 
particular choice of PDF as long as one restrict itself to the more recent 
sets.

Notwithstanding the differences highlighted in the previous section 
regarding the analytic formula of the matrix element and of the total width, 
the numerical value of the total cross--section 
is within \( \approx 20 \%\)  of that reported 
in~\cite{almeida} as can be seen in Fig.~\ref{fig3}: this can be 
easily understood as the cross--section could be written, 
to a good approximation, as a product 
of the production cross--section $\sigma(pp{\to}{\ell}^{+}N$) 
(independent from $\Gamma_N$), times the branching ratio 
\(\Gamma(N{\to}{\ell}^{+}W^{-})/\Gamma_N\) that is numerically 
the same although the formulas are different. 

Differences are however found on angular 
distributions of decay products (leptons and/or jets), 
that may be very important for the detection 
of the signal. Notice that from theoretical arguments~\cite{petcov} it can be 
stated that a forward--backward symmetry should show up in processes of 
production of heavy Majorana through CP conserving interactions like that 
here considered.   
Figs.~\ref{fig4}\&\ref{fig5} show normalized angular 
distributions of the outgoing lepton(s) and  gauge boson. 
Let \({\Theta}_{\ell}\) (\({\Theta}_{W}\)) be the angle formed by 
\(\vec{p}_{\ell}\) (\(\vec{p}_{W}\)) with the positive direction of the 
collision axis, in the center of mass system (c.m.s.) of the hadronic (\(pp\))
reaction (lab frame). 
Figs.~\ref{fig4}\&\ref{fig5} show 
the expected forward-backward symmetric behavior 
and in particular that 
the final state particles are emitted preferentially 
parallel or anti-parallel to the 
collision axis. 
Fig.~\ref{fig6}  shows the normalized distribution in the opening angle of 
the two like sign leptons.  
Here a difference is found relative to 
the numerical results reported in ref.~\cite{almeida}. 
In Fig.~\ref{fig6} one can observe that for relatively low neutrino masses 
the two leptons  are emitted  preferentially parallel to each 
other while at high neutrino 
masses  the distribution tends to become flatter.

The opposite behaviour is instead predicted in ref.~\cite{almeida} (see
their Fig.~5). 
This can again be 
explained assuming the interpretaion given above of the calculation reported
in~\cite{almeida}. The absence of the ghost particles amounts to the absence 
of the longitudinal polarization of the W-gauge boson (\(J^W_z=0\)): 
so the W gauge boson would have only the two helicity states  
\(\lambda=\pm 1\).
Suppose  a head-on collision between a \(u \) and  \(\bar{d}\) with 
\(\vec{p}_u\) along  the positive direction of the \(z\) axis. As $u$ has 
$\lambda=-\frac{1}{2}$ and $\bar{d}$ has $\lambda=+\frac{1}{2}$, the initial 
total angular momentum is $J_z = -1$. 
To conserve angular momentum along the \(z\)
axis, 
the only allowed configuration in the final state, composed of the two identical leptons and an on-shell \(W\),  is with a W-gauge boson with $\lambda=-1$ and the two positrons 
emitted with opposite momentum (antiparallel), since the two leptons have the same helicity, 
to give \(J^{\ell\ell}_z=0\). If we start with \(\bar{d} \; u\) we have an 
initial $J_z=+1$, so the final \(W\) has $\lambda=+1$ and the two 
leptons are again emitted antiparallel. However in reality the W-gauge 
boson has also the longitudinal 
polarization and the \(J^W_z=0\) possibility is also present: 
when this  is so, the two leptons must be parallel to give 
\(J^{\ell\ell}_z=\pm 1\). This additional helicity amplitude prevents
the angular distribution to drop as \(\theta_{\ell\ell}\to 0\). 

The other distributions discussed in ref.~\cite{almeida} 
(c.f.~transverse W-mass, transverse W-momentum   
and invariant mass  \(M_{\ell W},M_{\ell\ell W}\))
are confirmed by the present calculations and so are not repeated 
in this report.      
Fig.~\ref{fig7} shows the angular distribution with respect to the opening
angle of the jets in the fourth order process.
It is worth noticing that the 
W-angular-distribution, Fig.~\ref{fig5}, is more strongly 
peaked along the collision axis than that of the leptons, Fig.~\ref{fig4}. 
Thus the jets that originate from its decay are expected to be 
emitted preferentially along  the collision axis
and parallel to each other. This is indeed confirmed by the parton level
simulation as shown in Fig.~\ref{fig7}.   
If one takes into account that in the 
hadronization process more than two jets may be present, it is clear that the 
mass reconstruction of the W may not be so easy from the experimental point
of view. 
 
The peculiar characteristics of the signal are: two like sign leptons with the 
same angular and transverse momentum distributions, no missing energy, and at 
least two jets with an invariant mass distribution peaked 
at the W mass while  the invariant mass distributions in the  \( jj\ell \)
system will be peaked at the heavy neutrino mass, as shown in~\cite{almeida}. 
Clearly no SM process have these features but some have final states 
with accompanying neutrinos (missing energy) which are a source of potential 
background to the LSD + 2jets signal. 
All previously  cited literature  considered only 
heavy quark ($t\overline{t}$, $b\overline{b}$) production and subsequent 
decay as a major source. 
However recent simulations~\cite{dreiner} have recognized 
that more dangerous background at LHC can be single top production and  
gauge boson pair production (WZ and ZZ) followed by leptonic decays. After 
all cuts (for details see ref.~\cite{dreiner}) the authors find a 
background of $4.9\pm 1.6$ events with an integrated 
luminosity of 10 fb $^{-1}$ that 
implies a total cross \({\sigma}\, {\simeq}\, 0.5\) fb. This would  
restrict the observability  of the signal studied here 
in the neutrino mass range $100-250$ GeV. However the kinematic cuts used 
here are softer  than those used for the SM background suppression in  
ref.~\cite{dreiner}, so it is interesting to understand if the conclusions 
above can be significantly affected by changing the kinematic cuts. 
In~\cite{dreiner} is indicated that at 
least a cut of \(40\) 
GeV on \(p_T^{lepton}\) is necessary to suppress the SM 
background (also an isolation cut is needed but this will not affect the 
signal significantly). In Fig.~\ref{fig3} the dashed curve shows 
the effect on the
signal of increasing  the cut on the \(p_T^{lepton}\) up to \(40\) GeV: 
the total cross--section is lowered by a factor of up to an order of
magnitude in the  region of low neutrino masses, but it 
is practically unchanged  in the high mass region. 
The  drastic change in shape  found at small masses can be qualitatively 
understood by noticing that when   
\(M_N\) is small a larger part of the energy goes into  
\(M_W\) mass leaving not much kinetic 
energy for the outgoing  particles, so a high cut on \(p_T\) naturally 
lowers the cross--section. This is in agreement with the lepton 
transverse momentum distributions (which are not reported here as 
they are the same of those reported in ref.~\cite{almeida}): 
these are peaked at 
\(p_T^{lepton} \approx M_N/2 \). Clearly a cut of \(40 \) GeV on 
\(p_T^{lepton}\) cuts away most of the events  when \(M_N \approx 100\) GeV, thus causing a drastic change in \(\sigma_{tot}\).
However it should be remarked that 
also within these strong cuts there still remain a measurable 
number of events.

\section{Conclusions}
This paper has addressed the lepton number violating processes \((i)\) 
\(pp\, {\to}\,  \ell^{+}\ell^{+}\, +\, 2\hbox{jets}\) and \((ii)\)
\(pp\, {\to}\, \ell^{+}\ell^{+}W^{-}\) within the scenario of 
models where heavy Majorana neutrinos mix with light leptons. 
In particular regarding the fourth order process \((i)\) explicit 
analytic expressions of the helicity amplitudes 
are provided for the two competing mechanisms:  the WW-fusion and  
the resonant neutrino production. These helicity amplitudes 
might also be useful for more realistic 
(detector dedicated) Monte Carlo simulations and/or within the scenario of 
other models beyond the standard electroweak theory.
With respect to process \((ii)\) a detailed study, both analytical 
and numerical, with a thorough comparison  to the work 
of ref.~\cite{almeida}, has been performed
finding some differences in the analytical formula of the matrix element
which have been explained by noticing that the formula given in 
ref.~\cite{almeida} does not include the ghost contribution needed 
in the Feynman-'t Hooft gauge. However, as important as this might be, 
the main conclusion of ref.~\cite{almeida} (the size of \(\sigma_{tot}\)) 
is only affected by an error of \( \approx 20\% \),  though relevant  
differences are found on some of the angular
distributions. 

In particular the distribution in the opening angle (\(\Theta_{\ell\ell}\))
of the two leptons (Fig.~\ref{fig6}) is found to be significantly different  
from that reported in~\cite{almeida}.  

According to the study presented here 
in the framework of the mixing model
of heavy left--handed neutrinos there is a mass range of up 
to \(M_N \approx 800 \) GeV where the cross--section 
would be in principle observable ( \( 1 \)  event per year
assuming an integrated luminosity of \( 100 \) fb\(^{-1}\)). 
This window in the heavy neutrino masses is drastically reduced 
down to \( M_N \approx 250 \) GeV taking into account the recent 
SM background estimates  
quoted  in ref.~\cite{dreiner} 
(\(\approx  50 \pm 16\) events/year with an integrated 
luminosity  of \( 100 \) fb\(^{-1}\)), 
which appears to be dominated  by the production of 
double gauge bosons and single top.

The observation
of the LSD signal, in this model, is thus ultimately related to
the size of the mixing angle and the possibility of suppressing 
the SM background.
If a like sign 
di-lepton signal will be seen at LHC, then the question of 
determining the underlying physics scenario,
that triggers the non SM signal, would immediately arise;
more realistic,  detector dedicated, Monte Carlo
simulations would be needed and, of course,  
this eventuality would require additional theoretical
work. It is the authors' belief that the present work would in any case
be of importance in this direction.

\begin{acknowledgments}
C.~C. acknowledges support of an INFN grant (PG-102061/2000)
that allowed his stay in Perugia.
\end{acknowledgments}

\appendix
\section*{Total decay width of the heavy neutrino}

The expression for the partial decay widths of the heavy neutrino states
as deduced from the interaction Lagrangian given in Eq.~(\ref{leff})
are given by:
\begin{eqnarray}
\label{gammawnl}
\Gamma(N_\ell \to W \ell)&=&\frac{\alpha}{16\,s_W}\, \frac{M_N^3}{M_W^2}\,
|B_{eN}|^{2} \, \Phi\left( x_W \right),\\
\label{gammaznl}
\Gamma(N_\ell \to Z {\nu}_{\ell})&=&\frac{\alpha}{16\, s_W}\,
\frac{M_N^3}{M_W^2} \,
|B_{eN}|^2\, |B_{\nu N}|^2
\, \Phi\left( x_Z\right),  
\end{eqnarray}
with \(\Phi(x)= (1-x)^2 (1+2x)\), \( x_{W,Z} ={M_{W,Z}^2}/{M_N^2}\) and 
\(s_W=\sin^{2}\theta_W\), where \(\theta_W\) is the Weinberg angle, $\alpha$
the fine structure constant and $|B_{\nu N}|^2$ is determined from 
Eqs.~(\ref{effmix}). 

The total decay width is given by
\begin{equation}
\label{width}
\Gamma_N =
2\, \Gamma(N_\ell \to W \ell) + \Gamma(N_\ell \to Z {\nu}_{\ell})
\end{equation}
The factor of two in Eq.~(\ref{width}) arises because, due to its
Majorana nature,
the heavy neutrino can decay into  $W^- \ell^+$ and $W^+ \ell^-$
{\em with equal probability}. In Table~I, the corresponding numerical
values are reported for the total decay width $\Gamma_N $.
Including Higgs interactions of the heavy Majorana states
will increase the neutrino total decay width $\Gamma_N$.

The formulas that reproduce the widths given in~\cite{almeida}, as 
explained in the text, are obtained using 't Hooft-Feynman gauge but 
{\em without} including the  Goldstone 
boson contribution, which amounts to use for the sum over the real 
W polarizations: \( \sum_\lambda \epsilon_\lambda(k) \epsilon^*_\lambda(k) 
= -\eta_{\mu\nu} \), i.e. that corresponding to a massless gauge boson. 
One finds:
\begin{eqnarray}
\label{gammawnl1}
\Gamma(N_\ell \to W \ell)&=&\frac{\alpha}{8\,s_W}\, {M_N}\,
|B_{eN}|^{2} \, \Pi\left( x_W \right),\\
\label{gammaznl1}
\Gamma(N_\ell \to Z {\nu}_{\ell})&=&\frac{\alpha}{8\, s_W\,c_W}\,
{M_N} \,
|B_{eN}|^2\, |B_{\nu N}|^2
\, \Pi\left( x_Z\right),  
\end{eqnarray}
with \(\Pi(x)= (1-x)^2 \) and \(c_W=\cos^{2}{\theta}_W\), 
which indeed reproduce the values of \(\Gamma_N\) given by \cite{almeida}
and reported in table I.

\begin{table}
\caption{Total decay width of the (electron) heavy neutrino 
$\Gamma_N = \Gamma(N \to W \ell) + \Gamma(N \to Z \nu)$, obtained through the
use of Eq.~(\ref{gammawnl}) and Eq.~(\ref{gammaznl}) with $B_{eN} = 0.0052$. }
\begin{tabular}{lcccccc}
$M_N (\hbox{GeV})$& 100&300&500&600&800&1000\cr
$\Gamma_N (\hbox{GeV})$&
$0.11\times 10^{-2}$&
$0.138\times 10^{+0}$&
$0.65\times 10^{+0}$&
$1.13\times 10^{+0}$&
$0.267\times 10^{+1}$&
$0.52\times 10^{+1}$\cr
$\Gamma_N (\hbox{GeV})$\tablenote{For comparison the values used by the 
authors of~\protect\cite{almeida} are also reported.}
&$0.650\times 10^{-3}$&-&-&-&$0.580\times 10^{-1}$&-\cr
\end{tabular}
\end{table}

\begin{figure}
\leavevmode
\caption{W fusion mechanism for production of Like-Sign-Dileptons 
         (LSD) in high e\-ner\-gy hadronic collisions. 
         These diagrams, including the appropriate exchanges, describe also
         quark quark scattering and anti-quark anti-quark scattering. 
        }
\label{wwfusion}
\end{figure}
\begin{figure}
\leavevmode
\caption{Production of LSD through quark anti-quark
         scattering. In addition to the diagram of Fig. 1 (virtual W fusion), 
         one must consider the $\ell^+ N_\ell $ production via  
         quark anti-quark annihilation with subsequent hadronic decay of the 
         heavy neutrino 
         $N_\ell \to \ell^+ q \bar{q}$ ($\ell = e, \mu, \tau$). In principle 
         for $q \bar q$ this process interferes with the one in Fig. 1.
         }
\label{qqbaranni}
\end{figure}
\begin{figure}
\leavevmode
\caption{
         Signal cross-section for \(pp\, {\to}\, \ell^+\;\ell^+\;W^-\)
         (\( \ell = e\)): 
         the solid line includes contributions from first and second 
	 generation quarks; the dot-dashed line includes 
         only the \(u\bar{d}\)   
         contribution which is clearly the dominant (\(90\%\)); 
         the dashed line 
         indicates the total cross--section when a higher 
         cut on the lepton transverse momentum \(p_T > 40\) GeV is applied.
         }
\label{fig3}
\end{figure}
\begin{figure}
\leavevmode
\caption {Lepton polar angle (\(\Theta_\ell\)) distribution in the third 
          order process \( p p \to e^+ e^+ W^- \) at LHC energy \( \sqrt{S}
          = 14\)  TeV, for different values of the heavy neutrino mass \(M_N\).
          \(\Theta_\ell\) is the {\em laboratory  polar angle} 
          i.e. the angle formed 
          by the lepton with the positive direction of the hadronic 
          collision axis. 
}
\label{fig4}
\end{figure}
\begin{figure}
\leavevmode
\caption{Distribution in the 
          gauge boson \(W^-\) polar angle (\(\Theta_W\)) 
         in the third 
         order process \( p p \to e^+ e^+ W^- \) at LHC energy \( \sqrt{S}
         = 14\)  TeV, for different values of the heavy neutrino mass \(M_N\).
         The same legend as in Fig~\ref{fig4} applies. The polar angle
         \(\Theta_W\) is defined in the labratory system i.e. with respect
         to the hadronic collision axis.
}
\label{fig5}
\end{figure}
\begin{figure}
\leavevmode
\caption{
         Distribution in  the opening angle between the two identical 
         leptons (\(\Theta_{\ell\ell}\)), defined in the laboratory frame,
         in the third 
         order process \( p p \to e^+ e^+ W^- \) at LHC energy \( \sqrt{S}
         = 14\)  TeV. The legend of Figure~\ref{fig4} applies.
        }
\label{fig6}
\end{figure}
\begin{figure}
\leavevmode
\caption{
         Fourth order process \( p p \to e^+ e^+ 2 jets \). 
         Distributions for the opening angle (in the laboratory frame) 
         between the two quarks (jets) 
         (\(\Theta_{jj}\)) coming from the \(W^-\) gauge boson 
         decay. The different curves refer to different values of the 
         heavy neutrino mass according to the legend of Figure~\ref{fig4}.
}
\label{fig7}
\end{figure}

\end{document}